\documentclass[final,5p,times,twocolumn]{elsarticle}

\usepackage{amsthm}
\usepackage{amssymb}
\usepackage[ruled]{algorithm2e}
\usepackage{amsmath}
\usepackage{subfigure}
\usepackage{amsthm}
\usepackage[usenames,dvipsnames]{color}

\usepackage{threeparttable}
\usepackage{textcomp}
\usepackage[T1]{fontenc}
\usepackage{booktabs}

\usepackage{multirow}
\usepackage{setspace}
\usepackage{stfloats}
\usepackage{fancyhdr}
\usepackage{geometry}
\biboptions{sort&compress}
\journal{Elsevier}
\biboptions{sort&compress}

\begin{document}

\begin{frontmatter}


\newcommand{\be}{\begin{equation}}
\newcommand{\ee}{\end{equation}}
\newcommand{\br}{{\mbox{\boldmath{$r$}}}}
\newcommand{\bp}{{\mbox{\boldmath{$p$}}}}
\newcommand{\bpi}{\mbox{\boldmath{ $\pi $}}}
\newcommand{\bn}{{\mbox{\boldmath{$n$}}}}
\newcommand{\balfa}{{\mbox{\boldmath{$\alpha$}}}}
\newcommand{\ba}{\mbox{\boldmath{$a $}}}
\newcommand{\bta}{\mbox{\boldmath{$\beta $}}}
\newcommand{\bg}{\mbox{\boldmath{$g $}}}
\newcommand{\bPsi}{\mbox{\boldmath{$\Psi $}}}
\newcommand{\bsigma}{\mbox{\boldmath{ $\Sigma $}}}
\newcommand{\bGamma}{{\bf \Gamma }}
\newcommand{\bA}{{\bf A }}
\newcommand{\bP}{{\bf P }}
\newcommand{\bX}{{\bf X }}
\newcommand{\bI}{{\bf I }}
\newcommand{\bR}{{\bf R }}
\newcommand{\bZ}{{\bf Z }}
\newcommand{\bz}{{\bf z }}
\newcommand{\bx}{{\mathbf{x}}}
\newcommand{\bM}{{\bf M}}
\newcommand{\bU}{{\bf U}}
\newcommand{\bD}{{\bf D}}
\newcommand{\bJ}{{\bf J}}
\newcommand{\bH}{{\bf H}}
\newcommand{\bK}{{\bf K}}
\newcommand{\bm}{{\bf m}}
\newcommand{\bN}{{\bf N}}
\newcommand{\bC}{{\bf C}}
\newcommand{\bL}{{\bf L}}
\newcommand{\bF}{{\bf F}}
\newcommand{\bv}{{\bf v}}
\newcommand{\bSigma}{{\bf \Sigma}}
\newcommand{\bS}{{\bf S}}
\newcommand{\bs}{{\bf s}}
\newcommand{\bO}{{\bf O}}
\newcommand{\bQ}{{\bf Q}}
\newcommand{\btr}{{\mbox{\boldmath{$tr$}}}}
\newcommand{\bNSCM}{{\bf NSCM}}
\newcommand{\barg}{{\bf arg}}
\newcommand{\bmax}{{\bf max}}
\newcommand{\test}{\mbox{$
\begin{array}{c}
\stackrel{ \stackrel{\textstyle H_1}{\textstyle >} } { \stackrel{\textstyle <}{\textstyle H_0} }
\end{array}
$}}
\newcommand{\tabincell}[2]{\begin{tabular}{@{}#1@{}}#2\end{tabular}}
\newtheorem{Def}{Definition}
\newtheorem{Pro}{Proposition}
\newtheorem{Exa}{Example}
\newtheorem{Rem}{Remark}
\title{Distributed Multi-sensor Multi-view Fusion based on Generalized Covariance Intersection}


\author{Guchong Li$^a$, Giorgio Battistelli$^b$, Wei Y$\text{i}^{*a}$, Lingjiang~Kong$^a$}

\address{$^a$School of Information and Communication Engineering, \\University of Electronic Science and Technology of China, Chengdu, 611731, China,\\ E-mail: guchong.li@hotmail.com; kussoyi@gmail.com; lingjiang.kong@gmail.com\\
$^b$Dipartimento di Ingegneria dell' Informazione, \\Universit\`a degli Studi di Firenze, Firenze, 50139, Italy, \\E-mail: giorgio.battistelli@unifi.it}

\begin{abstract}
Distributed multi-target tracking (DMTT) is addressed for sensors having different fields of view (FoVs).
The proposed approach is based on the idea of fusing the posterior Probability Hypotheses Densities (PHDs) generated by the sensors on the basis of the local measurements.
An efficient and robust distributed fusion algorithm combining the  \emph{Generalized Covariance Intersection} (GCI) rule with a suitable \emph{Clustering Algorithm} (CA) is proposed.
The CA is used to decompose each posterior PHD into well-separated components (clusters). For the commonly detected targets, an efficient
parallelized GCI fusion strategy is proposed and analyzed in terms of $L_1$ error.  For the remaining targets, a suitable compensation strategy is adopted
so as to counteract the GCI sensitivity to independent detections while reducing the occurrence of false targets. Detailed implementation steps using a Gaussian Mixture (GM) representation of the PHDs are provided.
Numerical experiments clearly confirms the effectiveness of the proposed CA-GCI fusion algorithm.
\end{abstract}
\end{frontmatter}

\section{Introduction}
\label{sec:intro}
Multitarget tracking (MTT) is a problem of great practical relevance in many different contexts, from traffic monitoring and defense to robotics.
In many cases, the information collected from multiple sensors has to combined so as to improve performance.
In this context, distributed MTT algorithms have recently gained a considerable attention because of their fault tolerance, flexibility, and reduced computational burden as compared to a centralized fusion framework
\cite{Chang1997,Mori2002,Marano2008,Schizas2007,Cattivelli2008}.
In the single-target case, many efficient distributed tracking solutions have been developed over the years both in linear and nonlinear settings. In the former case
distributed Kalman filters (KF) are usually adopted \cite{Cattivelli2018conf,Giorgio}, while in the latter case solutions are usually based either on nonlinear variant of the KF \cite{TAC} or on the particle filter (PF) \cite{DPF}.
In the multi-target case, the problem is substantially more challenging due to the many reasons, above all the unknown association between targets and measurements.
Traditional solutions  include Joint Probabilistic Data Association Filter (JPDAF) \cite{BarShalom} and Multiple Hypothesis Tracker (MHT) \cite{Streit1994,MHT}. However,
the JPDAF requires the number of targets to be known a priori while the MHT involves explicit association between targets and measurements, which can involve a large computational load.
An alternative solution is the so-called Probability Hypothesis Density (PHD) filter \cite{MahlerPHD,Mahler2000}, which is based on the theory of Random Finite Sets (RFSs) \cite{MahlerPHD,Mahlerbook}.
The PHD filter can avoid explicit data association and can deal with an unknown and time-varying number of targets.
The PHD filter is usually implemented by resorting to either Gaussian Mixtures (GMs) \cite{Vo2} or Sequential Monte Carlo (SMC) approximations \cite{VoSMC,Ristic2016}.

The state of the art approach to distributed multi-sensor MTT is \emph{Generalized Covariance Intersection} (GCI) \cite{MahlerGCI2000,ClarkGCI2010}, also called \emph{Exponential Mixture Density} (EMD) \cite{Dabak1992}.
The GCI fusion amounts to computing the density that minimizes the
 sum of the information gains (Kullback-Leibler divergences \cite{Hurley2002,Giorgio}, KLD) from local posteriors, thus avoiding the problem of double-counting of common information \cite{Battistelli1}.
In the past years, any distributed RFS filters based on the GCI fusion rule have been proposed \cite{Battistelli,Uney,Li,Battistelli1,SO-GMB,Fantacci2018}.

While the GCI fusion rule performs well in many settings, it has been observed that this fusion rule can be sensitive to a high miss-detection rate \cite{Wei_2017fusion}.
In fact, the GCI fusion rule tends to preserve only tracks that are present in all the local posteriors. This pitfall is exacerbated when the sensors have different fields of views (FoVs).
For this reason, recently, some techniques have been proposed for dealing with the problem of different FoVs within the GCI fusion framework.
For instance, two possible solutions are proposed  in \cite{Mark2017} for the SMC-PHD filter. In the first method, particles of different sensors are combined only when they are believed to represent the same target.
In the second method, particles corresponding to the same target are hierarchically clustered and used to perform state extraction.
However, both methods are prone to estimation errors and can  underestimate the target number when nearby targets appear.
A distributed fusion algorithm based on the SMC-PHD filter, abandoning the limitation of fully overlapping FoV, is proposed in \cite{Gan2016}, which classifies the received particles into common particle set and external particle set.
As for the GM-PHD filter,  \cite{Giorgio2} proposes a solution to handle different FoVs in the context of  simultaneous localization and mapping (SLAM). Specifically, the approach of \cite{Giorgio2}
is based on the idea of using a uniform intensity over the whole region of interest to initialize all the local PHDs, in order to model the total uncertainty on the target (landmark) positions in the unexplored region of the map.
A different solution was proposed in \cite{Vasic2016}, which amends the traditional GCI fusion algorithm by considering the distance between Gaussian components (GCs). However, this method cannot solve the false alarm problem and meanwhile the cardinality is overestimated.
Similar ideas have recently been exploited to extended the labeled-RFS filter to multi-view sensors \cite{Wang2018,Liu2017,Lisuqi2018}.

In this paper, we focus on the problem of distributed multi sensor MTT  using the GM-PHD filter for a sensor network with different FoVs.
An efficient fusion strategy is developed that makes use of a clustering algorithm (CA) to improve robustness and performance of GCI fusion rule.

The main contributions of this paper are summarized as follows.
\begin{enumerate}
  \item We analyse the GCI fusion mismatch phenomenon caused by the limited sensor FoVs both from the theoretical point of view and via a simulation experiment.
  \item Motivated by the aforementioned analysis, we propose an effective and robust solution for GCI fusion of PHDs generated from sensors with different FoVs that is immune to the effect of fusion mismatch.
  The solution includes two parts: a parallelized GCI fusion in the commong FoV, and a compensation strategy outside the common FoV.
  \item We provide an error analysis in terms of  $L_1$-norm between the parallelized GCI fusion and the traditional GCI fusion algorithm. Specifically, we show that the error tends to zero
  when the partial intensities generated by the clustering algorithm are well separated.
  \item {We implement the proposed fusion algorithm based on the GM-PHD filter with adaptive birth model.} In the implementation, the \emph{union find set} (UFS) \cite{Ying2007,Pan2009} algorithm
  is adopted to ensure that the GC subsets generated during clastering are disjoint, and \emph{Optimal SubPattern Assignment} (OSPA) metric \cite{Schumacher} is used to compute the similarity between different subsets. In addition, \emph{Murty's algorithm} \cite{Murty1968} is utilized to obtain the association pairs effectively.
\end{enumerate}

The remainder of the paper is organized as follows. Section II introduces the main background notions including
the multi-target Bayes filter, the PHD filter, and the GCI fusion rule.
The mismatch of the GCI fusion for multi-view sensors is analyzed in Section III. The proposed fusion algorithm is described in Section IV,
 including a detailed description of the proposed GM implementation. Section V is dedicated to the simulations, while
 conclusions and perspectives for future work are given in Section VI.

\begin{table}
{
  \begin{center}
     \caption{Notations} \label{table11}
\hrulefill
\begin{itemize}
  \item Small letters, e.g., ${\mathbf{x}}$ and $z$, respectively, denote single-target states and single-target observations.
  \item Capital letters, e.g., ${\mathbf{X}}$ and $\mathbf{Z}$, respectively, denote multi-target states and multi-target observations.
  \item Blackboard bold letters, e.g., $\mathbb{X}$ and $\mathbb{Z}$, respectively, denote the state and measurement spaces.
  \item Each single target state $ {\mathbf{x}} = [ {{\mathbf{p}}}^\top $ ${{\mathbf{v}}}^\top ]^\top \in \mathbb{R}^{2 v} $, comprises position ${\mathbf{p}}_k^i \in \mathbb{R}^{v}$ and velocity ${\mathbf{v}}_k^i \in \mathbb{R}^{v}$, where `$\top$' denotes transpose.
  \item The multi-target state at time $k$ can be represented as
      \begin{equation}
      \nonumber
      \begin{aligned}
      {\mathbf{X}}_k &=& \{{{\mathbf{x}}_k^{1}, {\mathbf{x}}_k^{2},..., {\mathbf{x}}_k^{n_k} }\} 
      \end{aligned}
      \end{equation}
      where ${\mathbf{x}}_k^{i}$ is the state of the $i$-th target and $n_k$ is the number of targets.
  \item The multi-target observations from sensor $l$ at time $k$ can be represented as
      \begin{equation}
      \nonumber
      \begin{aligned}
      {\mathbf{Z}}_k^l &=& \{z_k^{l,1},z_k^{l,2},...,z_k^{l,m^l_k}\} 
      \end{aligned}
      \end{equation}
      where $z_k^{l,j}$ is the $j$-th observation data from sensor $l$ and $m_k^l$ is the number of observations.
  \item For a given target state set ${\mathbf{X}}_k$, each target ${{\mathbf{x}}_k^{i}} \in {\mathbf{X}}_k$ either survives at time $k+1$ with  survival probability $p_S\left( {\mathbf{x}}_k^{i} \right)$ or disappear with  probability $1-p_S\left( {\mathbf{x}}_k^{i} \right)$.
  \item A target having state $\mathbf{x}$ is detected by sensor $l$ with probability  ${p_D^l(\mathbf{x}})$.
\end{itemize}
\hrulefill
\end{center}
}
\end{table}
\section{Background}
\label{sec:models}
{This section provides an overview of the background material relevant to our work.
Some of the related notations are summarized in Table I.}
\subsection{Multi-target Bayes filter}

{ In this paper, the multi-target state is modelled as a RFS, i.e. a variable which is random in both the number of elements
(i.e. the number of targets) and the values of the elements (i.e. the target states).
From a probabilistic point of view,
the RFS $ {\mathbf{X}}_k$ is completely characterized by the multi-object posterior ${\pi_{k}}\left( \mathbf{X}_k \right) $,
corresponding to the multi-object density of $ {\mathbf{X}}_k$ conditional to all the observations collected up to and including time $k$.}

{ Given the multi-target posterior at time $k-1$ and the observation set $\mathbf{Z}_k$ at time $k$, the multi-target Bayes filter propagates the multi-target posterior in time via the Chapman-Kolmogorov prediction
\begin{equation}\label{eq:MTB1}
\begin{split}
&{\pi_{k|k - 1}} \left( \mathbf{X}_k \right) = \int {{f_{k|k - 1}}({\mathbf{X}_k}|{\mathbf{X}_{k - 1}}){\pi_{k - 1}}({\mathbf{X}_{k - 1}})\delta {\mathbf{X}_{k - 1}}}
\end{split}
\end{equation}
and then updates the density using the Bayes formula
\begin{equation}\label{eq:MTB2}
{\pi_{k}}\left( {\mathbf{X}_k} \right){\rm{ = }}
\frac{
g_{k} \left( {{\mathbf{Z}_k}|{\mathbf{X}_k}} \right) \pi_{k|k - 1} \left( \mathbf{X}_k \right)}{{\int {{g_{k}}\left( {{\mathbf{Z}_k}|{\mathbf{X}_k}} \right){\pi_{k|k - 1}} \left( \mathbf{X}_k \right)\delta {\mathbf{X}_k}} }}
\end{equation}
where $f_{k|k-1}(\mathbf{X}_{k}|\mathbf{X}_{k-1})$ and ${g_{k}}( {{\mathbf{Z}_k}|{\mathbf{X}_k}})$ are the multi-target transition density and the multi-target likelihood, respectively. The integrals
in (\ref{eq:MTB1})-(\ref{eq:MTB2})
are to be intended in the set-integral sense \cite{Mahlerbook}.
}

\subsection{The PHD filter}
{
The multi-target Bayes filter described in the previous section is in general intractable from the computational point of view.
A possible approximation is the so-called PHD filter \cite{MahlerPHD,Mahler2000} that is based on the idea of representing the multi-target set as a Poisson RFS.
The multi-object density $\pi ( {\mathbf{X}}) $ of a Poisson RFS $ {\mathbf{X}}$ takes the form
\begin{equation}
\label{Poisson1}
{\pi}(\mathbf{X}) = {\exp} \left ( {- \int_{\mathbb X} v(\mathbf{x} ) \, d \mathbf{x}} \right ) \,\prod\limits_{\mathbf{x} \in \mathbf{X}} {{v}(\mathbf{x})}
\end{equation}
where ${v}(\mathbf{x})$ is the so-called {\em PHD} or {\em intensity function}. The PHD ${v}(\mathbf{x})$ can be interpreted as the local density of the number of targets
in the single-target state space. In fact, given a region $\mathcal X \subseteq \mathbb X$,
the expected number of targets inside such a region can be computed as
$
 \int_{\mathcal X} v(\mathbf{x} ) \, d \mathbf{x} \, ,
$
and the total number of expected targets in the whole state space is given immediately by
$
 \int_{\mathbb X} v(\mathbf{x} ) \, d \mathbf{x} \, .
$
The PHD filter propagates in time the PHD ${v_k}(\mathbf{x})$ of the multi-target state conditioned to all the collected observations.
The detailed PHD filter recursions are given as follows.
\begin{equation}\label{eq:PHD_predict}
\begin{split}
{v_{k|k - 1}}(\mathbf{x}) = &\int {{p_{S}}(\zeta ){f_{k|k - 1}}(\mathbf{x}|\zeta ){v_{k - 1}}(\zeta )d\zeta }  + {\gamma _k}(\mathbf{x})
\end{split}
\end{equation}
\begin{equation}\label{eq:PHD_update}
\begin{split}
{v_k}(\mathbf{x}) = &[1 - {p_{D}}(\mathbf{x})]{v_{k|k - 1}}(\mathbf{x}) \\
&+ \sum\limits_{z \in {\mathbf{Z}_k}} {\frac{{{p_{D}}(\mathbf{x}){g_k}(z|\mathbf{x}){v_{k|k - 1}}(\mathbf{x})}}{{{\kappa _k}(z) + \int {{p_{D}}(\zeta ){g_k}(z|\zeta ){v_{k|k - 1}}(\zeta )d\zeta } }}}
\end{split}
\end{equation}
where $\gamma_k(\mathbf{x})$ is the prior PHD of target births at time $k$, and $\kappa_k(z)$ is the clutter density.
}

{
From (\ref{eq:PHD_predict}) and (\ref{eq:PHD_update}), we can see that the PHD filter operates in the single-target state space and avoids data associations.
In practice, instead of propagating the multi-target posterior density, the PHD filter
chooses to propagate the first-order moment of the true posterior by approximating the multi-target set as a Poisson RFS. The most typical implementations of the PHD filter are based either on GMs or SMC.
In this paper, we mainly focus on the GM implementation. Accordingly, the posterior PHD ${v_k}(\mathbf{x})$ is represented as
\begin{equation}\label{eq:GM}
{v_k}(\mathbf{x}) = \sum\limits_{i = 1}^{{N}} {\alpha _{k,i} \, {{\cal N}}(\mathbf{x};\mathbf{x}_{k,i},\mathbf{P}_{k,i})}
\end{equation}
where ${{\cal N}} (\mathbf{x};\mathbf{x}_{k,i},\mathbf{P}_{k,i})$ denotes the Gaussian PDF with mean $\mathbf{m}_{k,i}$ and covariance $\mathbf{P}_{k,i}$.
}

{
Concerning the birth model it can be either {\em predefined}  \cite{Vo2,VoSMC}  using prior information or {\em adaptive} \cite{Yan2015,Houssineau2011,Beard2012,Ristic2010,Ristic2012}.
The adaptive birth model (ABM) is based on the idea of using the measurements to initialize new components in the PHD, and is preferable when new targets can born in the whole state space.
As discussed in \cite{Yan2015,Houssineau2011,Beard2012,Ristic2010,Ristic2012}, the ABM makes use of a uniform (or partially uniform) birth intensity so as to reflect the uncertainty on the
location of newborn targets.
}

\subsection{GCI fusion rule}

{
Consider two multi-target posteriors $\pi^1_{k}\left({\mathbf{X}} \right)$ and $\pi^2_k\left({\mathbf{X}}\right)$ conditioned  on measurement set sequences from two different sensors.
When the correlation between the two measurement set sequences is unknown, the two multi-target posteriors can be fused by resorting to the so-called GCI fusion rule \cite{MahlerGCI2000}.
Under the GCI fusion rule, the two multi-target posteriors are fused into the multi-target posterior
\begin{equation}\label{eq: GCI_fusion}
\pi^{1,2}_k \left({\mathbf{X}} \right) = \frac{\pi^1_k\left({\mathbf{X}} \right)^{\omega_1}\pi^2_k\left({\mathbf{X}} \right)^{\omega_2}}{\int \pi^1_k \left({\mathbf{X}} \right)^{\omega_1}\pi^2_k \left({\mathbf{X}} \right)^{\omega_2}\delta {\mathbf{X}}}
\end{equation}
where the positive scalars $\omega_1,\omega_2$, with $\omega_1 + \omega_2 =1 $, are weights determining the relative importance of each multi-target posterior.
The fusion rule (\ref{eq: GCI_fusion}) was originally proposed  for the fusion of PDFs  in the statistics literature  \cite{genest1986combining}, and then generalizes to the case of RFS densities by Mahler \cite{MahlerGCI2000}.
As discussed in \cite{Heskes}, the fused density  given in (\ref{eq: GCI_fusion})  is the one minimizing the weighted sum of the KLD with respect to the densities to be fused
\begin{equation}\label{eq: GCI_fusion2}
{\pi ^{1,2}_k} = \arg \mathop {\inf }\limits_\pi  \left( {{\omega _1}{D_{KL}}(\left. \pi  \right\|{\pi ^1_k}) + {\omega _2}{D_{KL}}(\left. \pi  \right\|{\pi ^2_k})} \right)
\end{equation}
where
\begin{equation}
{D_{KL}}(\left. \pi  \right\|{\pi ^i}) \buildrel \Delta \over = \int {\pi (\mathbf{X})\log \frac{{\pi (\mathbf{X})}}{{{\pi ^i}(\mathbf{X})}}} \delta \mathbf{X}.
\end{equation}
See \cite{Battistelli}  for a proof in the case of RFS densities. While, for the sake of simplicity, only the case of two densities is considered, all the above considerations apply also in the case of more than two sensors.
}


{
Let now the multi-target posteriors to be fused be Poisson RFSs with PHDs $v^1_k(\mathbf{x})$ and $v^2_k(\mathbf{x})$, respectively, as it happens when each sensor runs locally a PHD filter.
Then, application of the GCI fusion rule yields again a Poisson RFS with fused PHD given by \cite{ClarkGCI2010}
\begin{equation}\label{eq:GCI}
{v^{1,2}_k }(\mathbf{x}) = \left [ {v^1_k}(\mathbf{x})\right ]^{\omega_1} \, \left [ {v^2_k}(\mathbf{x}) \right ]^{\omega_2}.
\end{equation}
}

{
While the result of the above cannot be computed in closed-form when the PHD is represented by a GM, because the exponentiation of a GM is in general not a GM, various approximation strategies exist.
For instance, one possible way of overcoming this drawback amounts to approximating each power $\left [ {v^l_k}(\mathbf{x})\right ]^{\omega_l}$ as a Gaussian mixture
\begin{equation}\label{eq:GM:approx}
\left [ {v^l_k}(\mathbf{x})\right ]^{\omega_l} \approx \sum\limits_{i = 1}^{{N^l}} {\widetilde \alpha _{k,i}^l \, {{\cal N}}(\mathbf{x}; \widetilde{\mathbf{x}}_{k,i}^l, \widetilde{\mathbf{P}}_{k,i}^l)}
\end{equation}
and, then, computing the fused PHD as
\begin{equation}\label{eq:GMPHD-GCI}
v^{1,2}_k (\mathbf{x}) = { {\sum\limits_{i = 1}^{{N^1}} {\sum\limits_{j = 1}^{{N^2}} {\alpha _{k,i,j}^{1,2} \, {\cal N}\left( {\mathbf{x};\mathbf{x}_{k,i,j}^{1,2} ,\mathbf{P}_{k,i,j}^{1,2}} \right)} } } }
\end{equation}
where
\begin{eqnarray}
{{\mathbf{P}}}_{k,i,j}^{1,2}
&=& {\left[ { \left (\widetilde {{\mathbf{P}}}_{k,i}^{1} \right)^{ - 1}} +{ \left (\widetilde{{\mathbf{P}}}_{k,j}^2 \right)^{ - 1}}\right]^{ - 1}}\\
{{\mathbf{m}}}_{k,i,j}^{1,2}
&=& {{\mathbf{P}}}_{k,i,j}^{1,2} \left [ {(\widetilde{{\mathbf{P}}}_{k,i}^{1})^{ - 1}}\widetilde{{\mathbf{x}}}_{k,i}^{1} + {(\widetilde {{\mathbf{P}}}_{k,j}^2)^{ - 1}}\widetilde {{\mathbf{x}}}_{k,j}^2 \right]\\
\label{GCI_weight}
\alpha _{k,i,j}^{12}
&=& \widetilde \alpha _{k,i}^1 \widetilde \alpha _{k,j}^2 \,
{\cal N}({\widetilde{\mathbf{x}}}_{k,i}^{1} - \widetilde{{\mathbf{x}}}_{k,j}^2;0,{{{\widetilde{\mathbf{P}}}_{k,i}^1}}+ {{{\widetilde{\mathbf{P}}}_{k,j}^2}}).
\end{eqnarray}
}

{
As discussed in \cite{Battistelli}, when the Gaussian Components (GCs) of each local PHD are well separated, the approximation in
(\ref{eq:GM:approx}) can be simply obtained as
\begin{eqnarray}
\widetilde{\mathbf{x}}_{k,i}^l & = & {\mathbf{x}}_{k,i}^l \\
\widetilde{\mathbf{P}}_{k,i}^l & = & \frac{{\mathbf{P}}_{k,i}^l}{\omega_l }\\
\widetilde \alpha _{k,i}^l &=& \left ( \alpha _{k,i}^l \right )^{\omega_l} \, \kappa (\omega_l ,{{\mathbf{P}}}_{k,i}^l)
\end{eqnarray}
where
\begin{equation}
\kappa\left( {\omega,\mathbf{P}} \right) \approx \sqrt {\frac{{\left| {2\pi \frac{{{\mathbf{P}}}}{\omega}} \right|}}{{{{\left| {2\pi \mathbf{P}} \right|}^\omega }}}} \; .
\end{equation}
More accurate approximations, like the one of \cite{Gunay2017Chernoff} based on sigma-points, must instead be used in the case of closely-spaced Gaussian components.
}

{
It can be seen from (\ref{eq:GMPHD-GCI}) that,
there will be $N_1N_2$ GCs after fusion. In the general case of $S$ sensors,
centralized GCI fusion \cite{Liguchong2} would yield $\prod\nolimits_s {{N_s}}$ GCs, thus requiring a high computational load.
Hence, when $S > 2$ a sequential fusion strategy \cite{Meng2009} may be preferable in practice.
}

\section{Analysis of the mismatch of GCI fusion for sensors with different FoVs}

{In this section, we provide a brief overview of the behavior of GCI fusion when applied to sensors with different FoVs, and we discuss the reason why GCI fusion may fail in this case.
To this end, we make use of the following  example.}

\begin{figure}[!htbp]
\centering
\begin{minipage}[b]{0.48\linewidth}
  \centerline{\includegraphics[width=1.1\columnwidth,draft=false]{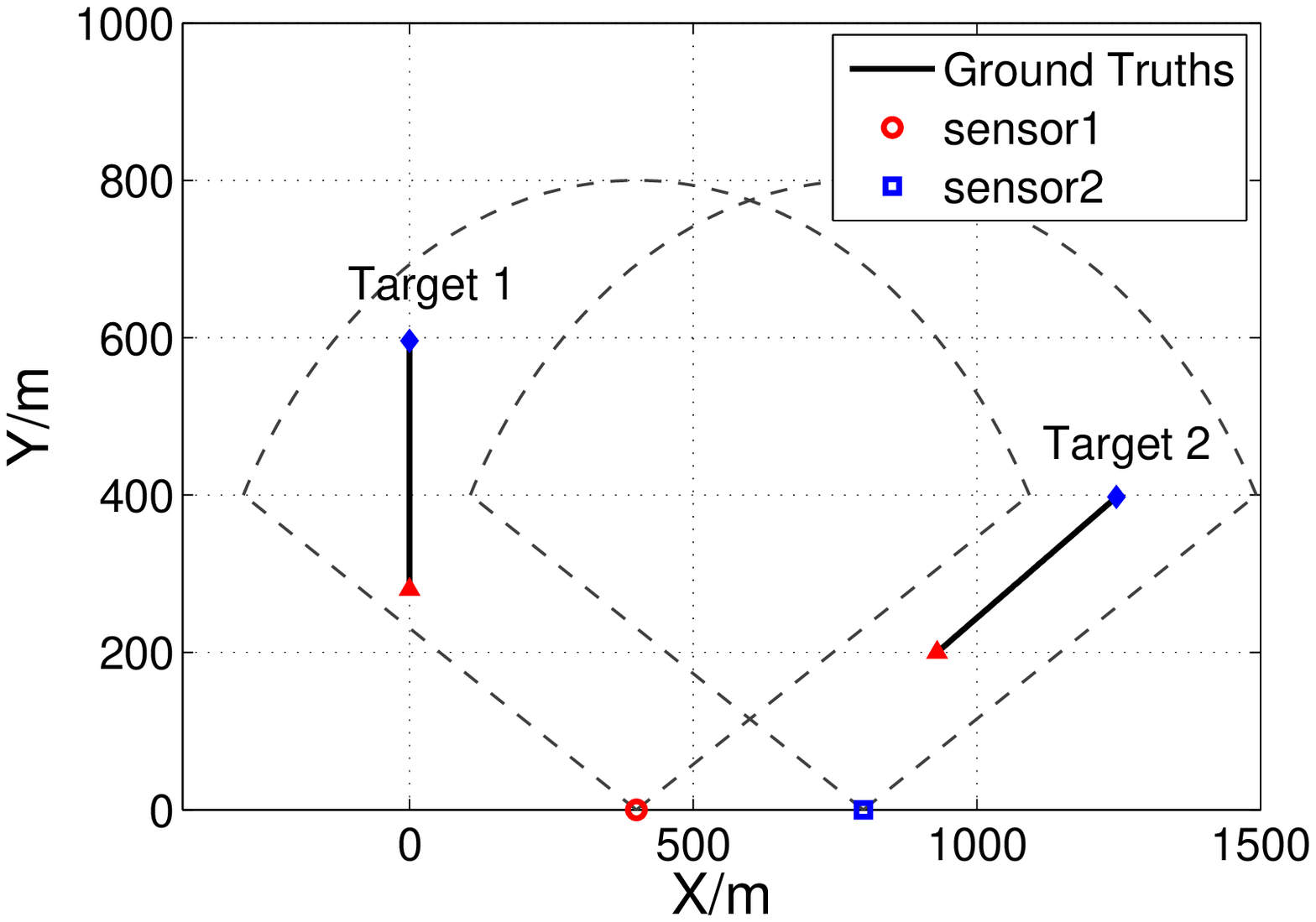}}
  \centerline{(a)}\medskip
\end{minipage}
\hfill
\begin{minipage}[b]{0.48\linewidth}
  \centering
  \centerline{\includegraphics[width=1.05\columnwidth,draft=false]{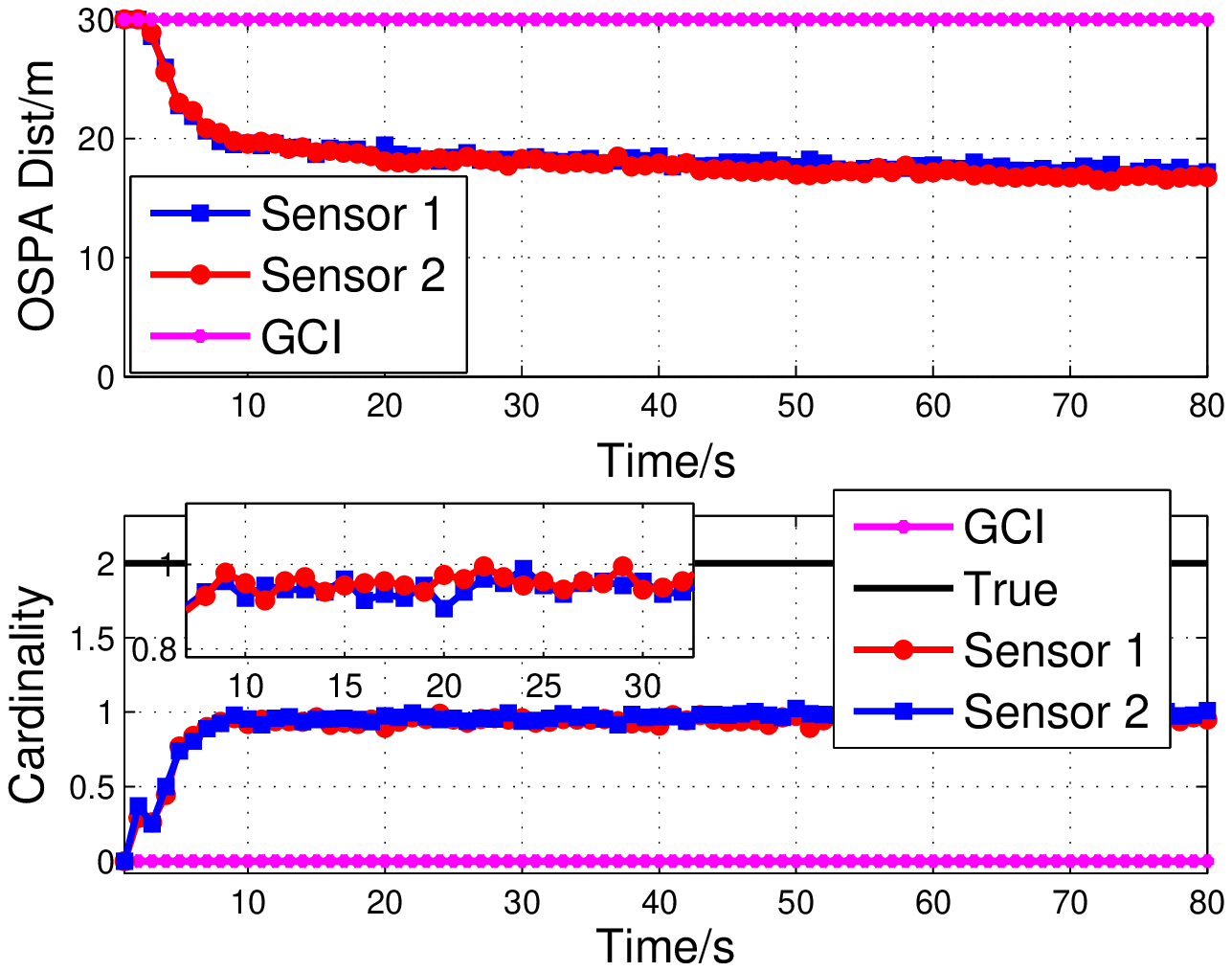}}
  \centerline{(b) }\medskip
\end{minipage}
\caption{(a) Simulation scenario used in Example 1; (b) performance of single-sensor PHD filters and standard GCI fusion in terms of OSPA errors and cardinality estimates.}
\label{fig:drawbacks}
\end{figure}

{
\emph{Example 1:} Consider a distributed sensor network with two sensors, each one running a GM-PHD filter with adaptive birth model (AB-GMPHD). The surveillance region is $[0,1500] m \times [0,1000] m$. The dynamics and observation models are the same as in \cite{CPHD}. The survival probability is $p_{S,k}=0.99$. The positions of the two sensors are
\begin{equation}
\nonumber
{p_1} = {\left[ {400,0} \right]^{\top}},{p_2} = {\left[ {800,0} \right]^{\top}}
\end{equation}
Each sensor has a limited FoV and, specifically, can detect only targets with relative angle in the interval $[-60^\circ,60^\circ]$.
The detection probability  is $p_{D}=0.98$ within the FoV, and zero otherwise.  The true tracks are shown in Fig. \ref{fig:drawbacks} (a) and the simulation results including OSPA errors \cite{Schumacher} and cardinality estimates are shown in Fig. \ref{fig:drawbacks} (b). The OSPA parameters are set equal to $c = 30 m$, $p = 2$.
}

{ It can be seen from Fig. \ref{fig:drawbacks} (a) that target 1 is detected only by sensor 1, while target 2 is detected only by sensor 2.
Accordingly, the PHD of sensor 1 will not contain GCs representing target 2, while the PHD of sensor 2 will not contain GCs representing target 1. Hence, the GCs of sensor 1 will be far from the GCs of sensor 2.
The result is that tracking performance after fusion is substantially worse than that of a single sensor.
}

{The reason for the behavior highlighted in Example 1 is that the essence of GCI fusion rule is the weighted multiplication between target densities, and the fusion process will work well only when both sensors detect the same targets. In fact, when a sensor does not detect a target, its PHD will be close to zero in the corresponding region of the state space. Hence, even if the other sensor is able to detect the target and, hence, has GCs with no-negligible weights in that region, application of the GCI fusion rule will yield a substantial decrease of such weights. Consequently, the target may be lost in the fused multi-target distribution, as it happens in Example 1.
This state of affairs can be also understood by looking at the GM-based implementation (\ref{eq:GMPHD-GCI}) of GCI fusion. In fact, we can see from (\ref{GCI_weight})
that a large distance between GCs results in a small fused weight, since ${\cal N}({\widetilde{\mathbf{x}}}_{k,i}^{1} - {\widetilde{\mathbf{x}}}_{k,j}^2;0,{{{\widetilde{\mathbf{P}}}_{k,i}^1}} + {{{\widetilde{\mathbf{P}}}_{k,j}^2}})$ tends to $0$
when the distance ${\widetilde{\mathbf{x}}}_{k,i}^{1} - {\widetilde{\mathbf{x}}}_{k,j}^2 $ increases.
}.
\section{The CA-GCI fusion algorithm}

{In order to overcome the drawbacks of the  traditional GCI fusion algorithm applied in multi-sensor fusion with partial overlapping FoVs,
we propose a novel clustering-based GCI fusion algorithm, named CA-GCI fusion algorithm.
First, we describe the basic ideas of the proposed fusion algorithm and analyse the approximation error as compared to traditional GCI fusion. Then, we discuss how the fusion rule can be modified so as to
deal with the different FoVs. Finally, detailed  guidelines on the implementation are provided.
}

{ For ease of presentation, the problem of fusing the PHDs coming from two sensors is considered.
All the presented techniques can be readily generalized to
sensor networks with multiple sensors ($S > 2$) by resorting to a sequential fusion strategy.}


\subsection{Fusion for overlapping FoVs}

{Let the two PHDs $v^1(\mathbf{x})$ and $v^2(\mathbf{x})$ to be fused be represented in GM form as in (\ref{eq:GM}).
Suppose now that the GCs of each of the two PHDs are grouped together in clusters so that the two PHDs can be decomposed as
\begin{eqnarray}
\label{subsets1}
{v^1}(\mathbf{x}) &=& \sum\limits_{p = 1}^{M_1} {\hat {v}_p^1(\mathbf{x})} \\
\label{subsets2}
{v^2}(\mathbf{x}) &=& \sum\limits_{p = 1}^{M_2} {\hat {v}_p^2(\mathbf{x})}
\end{eqnarray}
where each divided intensity function (DIF) ${\hat {v}_p^i(\mathbf{x})}$ consists of the linear combination of GCs.
The clusters are constructed so that:
\begin{enumerate}[(a)]
\item For each sensor $l$, clusters are well separated in the sense that for any $\mathbf{x} \in \mathbb X$ at most one of the
DIFs ${\hat {v}_p^l(\mathbf{x})}$ is substantially different from zero, i.e.,
\begin{equation}
 {\hat {v}_p^l(\mathbf{x})}  \, {\hat {v}_q^l(\mathbf{x})}  \approx 0
\end {equation}
for $l=1,2\,$ and for any $p \ne q$;
\item The first $Q$ DIFs of sensors 1 are matched with the first $Q$ DIFs of sensor 2.
These DIFs correspond to the regions of the state space in which both sensors have detected targets. Without loss of generality, we suppose that
${\hat {v}_p^1(\mathbf{x})}$ is matched with ${\hat {v}_p^2(\mathbf{x})}$, for $p = 1,\ldots,Q$, so that
\begin{equation}\label{DIF_b}
 {\hat {v}_p^1(\mathbf{x})}  \, {\hat {v}_q^2(\mathbf{x})}  \approx 0
\end {equation}
for any $p \ne q$;
\item The last $M_1-Q$ DIFs
of sensor 1 are not matched with any DIF of sensor 2, i.e.,
\begin{equation}
 {\hat {v}_p^1(\mathbf{x})} \, {\hat {v}_q^2(\mathbf{x})}  \approx 0
\end {equation}
for any $p = Q+1,\ldots,M_1$ and for any $q$. These DIFs correspond to targets detected only by sensor 1;
\item The last $M_2-Q$ DIFs
of sensor 2 are not matched with any DIF of sensor 1, i.e.,
\begin{equation}
 {\hat {v}_p^1(\mathbf{x})}  \, {\hat {v}_q^2(\mathbf{x})}  \approx 0
\end {equation}
for any $p $ and for any $q = Q+1,\ldots,M_2$. These DIFs  correspond to targets detected only by sensor 2.
\end{enumerate}
An efficient algorithm for the construction of the matched clusters based on the similarity between GCs will be provided later on in Section IV.D.
}

{
Given the matched clusters and the corresponding DIFs for the two PHDs, an approximate expression for the GCI fusion rule can be easily derived.
More specifically, in view of property (a) the DIFs are well separated and, hence, the exponentiations of the two PHDs can be approximated as
\begin{equation}
\left [ {v^l}(\mathbf{x})  \right ]^{\omega_l} =  \left [ \sum\limits_{p = 1}^{M_l} {\hat {v}_p^l(\mathbf{x})} \right ]^{\omega_i} \approx  \sum\limits_{p = 1}^{M_l} \left [ {\hat {v}_p^l(\mathbf{x})} \right ]^{\omega_l}
\end{equation}
for $l=1, 2 \,$. Further, by exploiting properties (b)-(d),  we also have that
\begin{equation}\label{matchedDIFs}
\left [ {v^1}(\mathbf{x})  \right ]^{\omega_1} \left [ {v^2}(\mathbf{x})  \right ]^{\omega_2}  \approx   \sum\limits_{p = 1}^{Q} \left [ {\hat {v}_p^1(\mathbf{x})} \right ]^{\omega_1}  \left [ {\hat {v}_p^2(\mathbf{x})} \right ]^{\omega_2} \, .
\end{equation}
This means that the matched DIFs can be fused in parallel and the fused PHD $v^{1,2}(\mathbf{x}) $ can be approximated by the intensity
\begin{equation}\label{eq:GCI_para}
\hat v^{1,2}(\mathbf{x}) = {{\sum\limits_{p = 1}^Q {{{\left[ {{\hat v}_p^1(\mathbf{x})} \right]}^{{\omega _1}}}{{\left[ {{\hat v}_{p}^2(\mathbf{x})} \right]}^{{\omega _2}}}} }}  \, .
\end{equation}
From the above equation, it is clear that only the DIFs corresponding to targets detected by both sensors are preserved after fusion, while the DIFs corresponding to targets detected by either sensor 1 or 2  (see points
(c) and (d) above) disappear. Hence, at a basic level, the GCI fusion rule can be interpreted as performing an {\em intersection} of the DIFs (and hence of the tracks) of the two sensors.
This is consistent with the behavior observed empirically in Section III. Hereafter, the approximated fusion rule (\ref{eq:GCI_para}) will be referred to as {\em parallelized GCI}.
\begin{Rem}
We point out that, in the worst case where all targets are so close that it is impossible to partition the PHDs into separated DIFs,  there will be only one DIF for each sensor
and the parallelized GCI fusion (\ref{eq:GCI_para}) will reduce to the standard GCI fusion algorithm.
\end{Rem}
}

\subsection{Analysis of the approximation error}

{
In this section, we analyze the error committed when the approximated fusion rule (\ref{eq:GCI_para}) is used.
To this end, it is convenient to partition the state space $\mathbb X$ into $M = M_1 + M_2 -Q$ subsets $\mathbb X_1$, $\ldots$, $\mathbb X_M$ so that
\begin{equation}\label{Subspace}
\mathbb{X} = \mathbb{X}_1 \cup \mathbb{X}_2 \cup \cdots \cup \mathbb{X}_{M}
\end{equation}
and
\begin{equation}
\mathbb{X}_p \cap \mathbb{X}_q = \emptyset
\end{equation}
for any $p \ne q$.
The partition is built so that:
\begin{enumerate}[(i)]
\item for $p = 1, \ldots,Q$, the set $\mathbb X_p$ contains the tracks represented by the matched DIFs $ {\hat {v}_p^1(\mathbf{x})} $ and $ {\hat {v}_p^2(\mathbf{x})} $
 in the sense that
\begin{equation}
  {\hat {v}_p^l (\mathbf{x})} \approx 0
\end{equation}
for $l=1,2$ and for any $\mathbf{x} \in \mathbb X \setminus \mathbb X_p$;
\item for $p = Q+1, \ldots,M_1$, the set $\mathbb X_p$ contains the tracks represented by the unmatched DIF $ {\hat {v}_p^1(\mathbf{x})} $ of sensor 1
 in the sense that
\begin{equation}
  {\hat {v}_p^1 (\mathbf{x})} \approx 0
\end{equation}
for $\mathbf{x} \in \mathbb X \setminus \mathbb X_p$;
\item for $p = M_1+1, \ldots, M$, the set $\mathbb X_p$ contains the tracks represented by the unmatched DIF $ {\hat {v}_{p-M_1+Q}^2(\mathbf{x})} $ of sensor 2
 in the sense that
\begin{equation}
  {\hat {v}_{p-M_1+Q}^2 (\mathbf{x})} \approx 0
\end{equation}
for $\mathbf{x} \in \mathbb X \setminus \mathbb X_p$.
\end{enumerate}
Notice that the above construction is possible in view of fact that the DIFs are supposed to be well separated and matched as discussed in points (a)-(d) above.
Notice also that this partition need not be actually determined, since it is just instrumental for the analysis, but it is not used in the actual fusion rule.
}

\begin{figure}
\begin{center}
\includegraphics[width=1.2\columnwidth,draft=false]{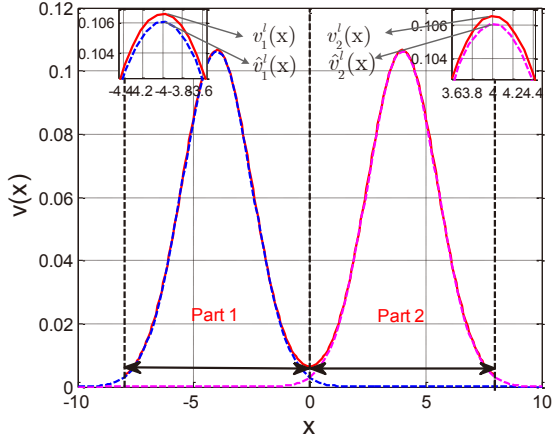}
\end{center}
\caption{The discrepancy between the approximated partial intensity and the true partial intensity for sensor $l$.}
\label{fig:subspace}
\end{figure}

{
Let us now decompose the two PHDs as
\begin{equation}\label{eq:partition_1}
{v^l}(\mathbf{x}) = \sum\limits_{p = 1}^{M} {v_p^l(\mathbf{x})} \, ,
\end{equation}
for $l=1,2$ ,  where each partial intensity $v_p^l(\mathbf{x})$ is defined as
\begin{equation}
v_p^l(\mathbf{x}) = v^l(\mathbf{x}) \, {\bf{1}}_{\mathbb{X}_p}(\mathbf{x})
\end{equation}
with $ {\bf{1}}_{\mathbb{X}_p}(\cdot)$ the {\em indicator function}
\begin{equation}
\bf{1}_{\mathbb{X}_p}(\mathbf{x}) =
\left\{
 \begin{array}{ll}
1 & \mbox{ if } \mathbf{x} \in \mathbb{X}_p \\
0 & \mbox{ otherwise}.
\end{array} \right.
\end{equation}
Notice that the partial intensities $v_p^l(\mathbf{x})$ satisfy properties (a)-(b) exactly in that
\begin{equation}
v_p^l(\mathbf{x}) \, v_q^r(\mathbf{x}) = 0
\end{equation}
for any $l,r $ and any $p \ne q$. As a consequence, the fused PHD $v^{1,2} (\mathbf{x})$ can be equivalently written as
\begin{equation}\label{eq:GCI_part}
v^{1,2}(\mathbf{x}) = {{\sum\limits_{p = 1}^M {{{\left[ {{ v}_p^1(\mathbf{x})} \right]}^{{\omega _1}}}{{\left[ {{ v}_{p}^2(\mathbf{x})} \right]}^{{\omega _2}}}} }} \, .
\end{equation}
Notice also that, since the DIFs are determined so as to approximately satisfy properties (a)-(d), we have that
\begin{equation}
\begin{array}{ll}
{{v}_p^1(\mathbf{x})} \approx { \hat {v}_p^1(\mathbf{x})} & \mbox{ for } p=1, \ldots, M_1 \vspace{1mm} \\
{{v}_p^1(\mathbf{x})} \approx 0 & \mbox{ for } p=M_1+1, \ldots, M \vspace{1mm} \\
{{v}_p^2(\mathbf{x})} \approx { \hat {v}_p^2(\mathbf{x})} & \mbox{ for } p=1, \ldots, Q \vspace{1mm} \\
{{v}_p^2(\mathbf{x})} \approx 0 & \mbox{ for } p=Q+1, \ldots, M_1 \vspace{1mm}  \\
{{v}_p^2(\mathbf{x})} \approx { \hat {v}_{p-M_1+Q}^2(\mathbf{x})} & \mbox{ for } p=M_1+1, \ldots, M
\end{array}
\end{equation}
The above relationships can be quantified by bounding the discrepancies in terms of a scalar $\delta$
\begin{equation}\label{bounding}
\begin{array}{rcll}
\| {{v}_p^1} - { \hat {v}_p^1} \|_\infty &\le& \delta  & \mbox{ for } p=1, \ldots, M_1 \vspace{1mm} \\
\| {{v}_p^1} \|_\infty &\le& \delta & \mbox{ for } p=M_1+1, \ldots, M \vspace{1mm} \\
\| {{v}_p^2} - { \hat {v}_p^2} \|_\infty &\le& \delta & \mbox{ for } p=1, \ldots, Q \vspace{1mm} \\
\| {{v}_p^2} \|_\infty &\le& \delta & \mbox{ for } p=Q+1, \ldots, M_1 \vspace{1mm}  \\
\| {{v}_p^2} - { \hat {v}_{p-M_1+Q}^2} \|_\infty &\le& \delta & \mbox{ for } p=M_1+1, \ldots, M
\end{array}
\end{equation}
where, given a function $f: \mathbb X \rightarrow \mathbb R$, $\| f \|_\infty$ denotes its infinite norm $\| f \|_\infty = \sup_{\mathbf{x} \in \mathbb X} | f(\mathbf{x})|$.
Clearly, the more well separated are the matched DIFs  ${\hat {v}_p^l(\mathbf{x})}$ the smaller is the scalar $\delta$. In other words, $\delta$ can be interpreted
as a quantitative measure of how much each DIF  ${ \hat {v}_p^l(\mathbf{x})}$ is concentrated in the corresponding domain of the partition. For the readers' convenience, Fig. \ref{fig:subspace} shows
the difference between the DIFs (in blue) and the partial intensities (in red) in a simple example.
}

{
The following result provides a quantification of the error committed when the parallelized GCI fusion
(\ref{eq:GCI_para})  is used in place of the exact GCI fusion (\ref{eq:GCI}).
\begin{Pro}
Let the two PHDs   ${\left[ {{v^1}(\mathbf{x})} \right]^{{\omega _1}}}$ and ${\left[ {{v^2}(\mathbf{x})} \right]^{{\omega _2}}}$ to be fused be bounded, and let
  \begin{equation}\label{bounded}
  K = \max \left\{ {{{\left\| {{{\left[ {v_p^1} \right]}^{{\omega _1}}}} \right\|}_1},{{\left\| {{{\left[ {v_q^2} \right]}^{{\omega _2}}}} \right\|}_1}} \right\}.
  \end{equation}
for any $p$, $q$, where $\| \cdot \|_1$ denotes  the $L_1$-norm, ${\left\| f \right\|_1} = \int {\left| {f(\mathbf{x})} \right|} d\mathbf{x}$. \\
Then, the following bound holds
 \begin{equation}
   {\left\| {\hat v^{1,2}- v^{1,2}} \right\|_{{1}}}  \le{M_2}K{\delta ^{{\omega _1}}} + \left( {2Q + {M_1}} \right)K{\delta ^{{\omega _2}}}.
  \end{equation}
\end{Pro}
The proof of Proposition 1 is given in the Appendix A. It follows from Proposition 1 that the discrepancy $ {\left\| {\hat v^{1,2}- v^{1,2}} \right\|_{{1}}} $ between
the parallelized GCI fusion (\ref{eq:GCI_para}) and the exact GCI fusion (\ref{eq:GCI}) is small when the scalar $\delta$ is small (i.e., when the matched DIFs are well separated).
}

\subsection{Fusion for different FoVs}

{When the sensors share the same FoV, the local PHDs contain information on the same targets and the GCI fusion performs well.
However, as discussed in Section III, when different sensors have different FoVs, the GCI fusion of local posteriors can lead to unsatisfactory
results. This is because, at a basic level, GCI fusion can be interpreted as performing an intersection among the tracks of different sensors.
In fact, recalling that GCI fusion of two PHDs can be approximated as (\ref{eq:GCI_para})), there will be $M_1-Q$ DIFs for sensor 1, namely $\hat v^1_{Q+1} (\mathbf{x}), \ldots, \hat v^1_{M_1} (\mathbf{x})$,
and $M_2-Q$ DIFs for sensor 2, namely $\hat v^2_{Q+1} (\mathbf{x}), \ldots, \hat v^2_{M_2} (\mathbf{x})$,
that do not participate in the fusion process so that the corresponding tracks disappear after fusion. Hereafter, we will refer to these DIFs as {\em unconfirmed DIFs}.
Then, the idea is to modify the fusion rule so as to preserve, at least partially, the information contained in these unconfirmed DIFs, in order account for the fact that each sensor has only a partial view of the overall surveillance area.
Of course, special care has to be taken in this operation since the unconfirmed DIFs, corresponding to tracks present in only one of the sensors, may actually be false targets.
With this respect, it is important to distinguish the unconfirmed DIFs contained in the intersection of the FoVs from those contained only in the FoV of one sensor.
In fact, while it is reasonable to think that the former ones correspond to false alarms, no similar conclusion can be drawn on the latter ones.
To this end, it is convenient to introduce the following concept.
\begin{Def}
Given an unconfirmed DIF $\hat v^l_p ( \mathbf{x} ) $, we say that $\hat v^l_p ( \mathbf{x} )$ is observed by sensor $r$ if
\begin{equation}\label{eq:threshold}
\int_{Fo{V_r}} {{{\hat v}_p^l}\left( \mathbf{x} \right) d \mathbf{x} > \gamma } \int_{\mathbb X} {{{\hat v}_p^l}\left( \mathbf{x} \right) d \mathbf{x} }
\end{equation}
where $\gamma \in (0,1)$ is a threshold.
\end{Def}
Let us now introduce, for each DIF of sensor $l$, a binary variable $\beta^{lr}_p$
taking value $1$ if $\hat v^l_p ( \mathbf{x} )$ is not observed by sensor $r$, and value $0$ otherwise.
Then, the idea is to preserve all the DIFs of one sensors that are not observed by the other.
Accordingly, we define the {\em preserved PHDs} of sensors $1$ and $2$ as
\begin{eqnarray}
\label{preservedPHD1}
{{\bar v}^1}(\mathbf{x}) &=& \sum_{p = Q + 1}^{M_1} \beta^{1 2}_p
\,  {\hat v_p^1(\mathbf{x})}\\
\label{preservedPHD2}
{{\bar v}^2}(\mathbf{x}) &=& \sum_{p = Q + 1}^{M_2}  \beta^{2 1}_p \, {\hat v_p^2(\mathbf{x})}.
\end{eqnarray}
}

{
We have now to distinguish two cases: namely complete trust, and partial trust.
The complete trust case refers to the situation in which the sensor FoVs are time-invariant, and
the sensors do not receive any feedback after fusion so that the local PHDs only contain information
on the targets detected within the local constant FoV. In this case, it is clear that it makes sense to perform fusion only in the intersection  $Fo{V_1} \cap Fo{V_2}$,
whereas the fusion center has to completely trust sensor $1$ in its exclusive field of view $Fo{V_1} \setminus Fo{V_2}$ (and, similarly, a complete trust should be given to
sensor $2$ in $Fo{V_2} \setminus Fo{V_1}$). Hence, in this case, the total fused intensity can be taken equal to
\begin{equation}\label{eq:trust}
\tilde v^{1,2} (\mathbf{x}) = \hat v^{1,2}(\mathbf{x}) + {{\bar v}^1}(\mathbf{x}) + {{\bar v}^2}(\mathbf{x}).
\end{equation}
}
{
On the other hand, in many situations, local PHDs also contain information outside the local FoV. For instance, this can happen when the FoV is time-varying,
like in the case of mobile sensors, or when there is a feedback from the fusion center to the local nodes\footnote{Even in a centralized setting, feedback is important.
In fact, thanks to the information fed back from the fusion center, each sensor node is able to keep track of all the targets currently present in the total surveillance area.
For instance, this means that when a target already detected by one sensor first enters the FoV of another sensor, no initialization is required because the predicted density
already contains a GC corresponding to that target.
}, or when the fusion is distributed/decentralized.
Clearly, in these cases, it would be inappropriate to completely trust a DIF that is present in one sensor but is not present in the other.
In this case, we follow  a different approach by introducing a {\em compensation intensity} in the PHDs to be fused
(a similar idea was proposed in \cite{Lisuqi2018} in the context of GCI fusion of labelled RFS densities).
This amounts to adding a new term to each local PHD that models the birth of new targets outside the current FoV.
The new term, which corresponds to a uniform (uninformative) intensity outside the current FoV, makes it possible  for each sensor to initialize new tracks from those of the other sensor
(similarly to what happens in adaptive birth model wherein a uniform birth density within the FoV is used to initialize new tracks from the measurements).
The value $\Delta > 0$ of this additional intensity plays the role of a confidence factor and can be also interpreted as the expected target intensity.
Then, the fused intensity can be written as
\begin{equation}\label{eq:no-trust}
\tilde v^{1,2} (\mathbf{x}) = \hat v^{1,2}(\mathbf{x}) + [{{\bar v}^1}(\mathbf{x})]^{\bar\omega_1} \Delta^{1-\bar\omega_1}+ [{{\bar v}^2}(\mathbf{x})]^{\bar\omega_2} \Delta^{1-\bar\omega_2}
\end{equation}
where the two additional terms arise from the fusion of the preserved PHD  ${{\bar v}^i}(\mathbf{x})$ of one sensor with the uniform (uninformative) intensity $\Delta$ of the other,
and $\bar\omega_1 \in \left[0,1\right]$ and $\bar\omega_2 \in \left[0,1\right]$ are fusion weights for non-overlapping areas, which can be different from $\omega_1$ and $\omega_2$.
Notice that $\Delta$, $\bar \omega_1$, $\bar \omega_2$ can be tuned according to the actual sensor network under consideration.
For instance, by setting $\Delta = 1$, $\bar \omega_1=1$ and $\bar \omega_2=1$ we retrieve the case of complete trust (\ref{eq:trust}), whereas by choosing $\Delta = 0$ we consider only the information
pertaining to the intersection of the FoVs.
}

\subsection{The implementation of the proposed algorithm}

{In this section,  the detailed GM implementation of the proposed CA-GMPHD-GCI algorithm is provided, including the construction of DIFs.
The GCI fusion between matched DIFs is efficiently  implemented in a parallelized way, while the preserved DIFs are fused  using the compensation intensity introduced in the previous section.}

{Given the PHDs of the two sensors in GM form (\ref{eq:GM}), the following eight implementation steps are carried out.}

{\underline{1) Grouping by pre-clustering:} Considering that the number of clusters is unknown, we choose the GCs whose weight $\alpha$ satisfies
\begin{equation}
\alpha > T_\alpha
\end{equation}
as candidate cluster centers, where $T_\alpha$ is a pre-specified threshold. Specifically, $T_\alpha$ is suitably tuned so as to ensure that newborn targets can be quickly detected while avoiding that the number of groups
becomes too large. The \emph{corrected Mahalanobis distance}
\begin{equation}\label{eq:M-distance}
 {\left( {{{\mathbf{x}}_1} - {{\mathbf{x}}_2}} \right)^{\top}}\left( {{{\mathbf{P}}_1}^{ - 1} + {{\mathbf{P}}_2}^{ - 1}} \right)\left( {{{\mathbf{x}}_1} - {{\mathbf{x}}_2}} \right)
\end{equation}
is used to measure the distance between two GCs having mean/covariance $({\mathbf{x}}_1, {\mathbf{P}}_1)$ and $({\mathbf{x}}_2, {\mathbf{P}}_2)$, respectively.
For each sensor, clusters are formed by grouping together all GCs whose distance from a cluster center is less than a given threshold $T_d$. The same distance criterion is used for the two sensors.
After the clustering process is finished, the GCs of the two sensors are decomposed into the following subsets
\begin{eqnarray}
{L^1} &=& \left\{ {L_1^1,L_2^1, \cdots ,L_a^1} \right\}\\
{L^2} &=& \left\{ {L_1^2,L_2^2, \cdots ,L_b^2} \right\}
\end{eqnarray}
where $a$ and $b$ are the numbers of subsets for sensor $1$ and $2$, respectively.
Due to the different FoVs, $a$ and $b$ are in general different.
}

{\underline{2) Making the clusters disjoint:}
The subsets resulting from the above-described procedure need not be disjoint, i.e. each GC is in general can be assigned to more than one subset $L_l^i$.
This usually happens when there are targets at close distance.
In order to ensure that each GC is associated with a single subset,  we resort to the \emph{union find set} (UFS) algorithm \cite{Ying2007,Pan2009}.
After application of UFS, a new decomposition is obtained for the GCs of the two sensors
\begin{eqnarray}
{L^1} &=& \left\{ {L_1^1,L_2^1, \cdots ,L_{M_1}^1} \right\}\\
{L^2} &=& \left\{ {L_1^2,L_2^2, \cdots ,L_{M_2}^2} \right\}
\end{eqnarray}
where $M_1 \le a$ and $M_2 \le b$ are the  new numbers of subsets for sensor $1$ and $2$, respectively, and where, by construction,
$L^l_p \cap L^l_q = \emptyset$ for any $p \ne q$.
Hence, denoting by $N^l_p$ the number of GCs in the subset  $L^l_p$ and by $N^l$ the total number of GCs of sensor $l$,
we have  ${N^l} = \sum\nolimits_{p  = 1}^{{M_l}} {N_p^l}$.
Notice that each of the subset $L^l_p$ defines a DIF $\hat v^l_p (\mathbf{x} )$ obtained as combination of the GCs in  $L^l_p$.
}


{\underline{3) Matching subsets:} }
{In order to find a matching between the subsets of sensor $1$ and those of sensor $2$, i.e. between the DIFs of sensor $1$ and those of sensor $2$,
we use the OSPA distance  \cite{Schumacher}  in order to measure the distance (dissimilarity) between subsets.
Then, the following ${M_1 \times M_2}$ distance matrix is constructed
\begin{equation}
\mathbf{D} = {\left[ {\begin{array}{*{20}{c}}
{{d_{1,1}}} \cdots {{d_{1,{M_2}}}}\\
 \vdots  \quad \ddots \quad \vdots \\
{{d_{M_1,1}}} \cdots {{d_{M_1,M_2}}}
\end{array}} \right]}
\end{equation}
where $d_{p,q}$ denotes the OSPA distance between $L_p^1$ and $L_q^2$.}

{Then an optimal matching between the subsets $L^1$ and $L^2$ is found by solving a linear assignment problem \cite{Kuhn}.
Specifically, Let us denote by $S$ the binary assignment matrix defined so that $S_{p,q} = 1$ if  $L_p^1$  is associated to $L_q^2$, and $S_{p,q} = 0$ otherwise.
Further let us suppose, without loss of generality, that $M_1 \ge M_2$.
Then, we use {\em Murty's algorithm} \cite{Murty1968} to find the matrix $S^*$ that minimizes
\begin{equation}
tr({S^{\top}}\mathbf{D}) = \sum\limits_{p = 1}^{M_1} {\sum\limits_{q = 1}^{M_2} {{d_{p,q}} \, {S_{p,q}}} }.
\end{equation}
under the constraint $\sum_{p=1}^{M_1}  S_{p,q} = 1$.
}

{
Afterwards, for every associated pair $(p,q)$ such that $S_{p,q} = 1$, we check whether the distance $d_{p,q}$ is below or above a given threshold $T_r$.
In the former case, the association is considered valid. Conversely, if the distance is too large, the association is discarded.
As a result, in general only $Q \le \min\{M_1,M_2\}$ pairs $(p,q)$ are considered valid associations (these are the DIFs corresponding to the targets detected by both sensors),
while $M_1-Q$ subsets of sensor $1$ and $M_2-Q$ subsets of sensor 2 are not associated
(these are the unconfirmed DIFs corresponding to the targets detected only by one of the sensors).
Finally, the subsets of the two sensors are reordered so that the first $Q$ pairs $(L_p^1,L_p^2)$, $p=1,\ldots,Q$, correspond to matched subsets/DIFs.
}

{\underline{5) Selecting the preserved DIFs:}
After the match procedure is finished, we resort to (\ref{eq:threshold}) to determine which of the unconfirmed DIFs should be preserved and which, instead, should be deleted.
Specifically, each unmatched subset $L^1_p$, $p = Q+1,\ldots,M_1$ of sensor is preserved (i.e. $\beta^{12}_p=1$) if and only if
\begin{equation}\label{GM_FTD}
\sum\nolimits_{i = 1}^{N_p^1} {\alpha _{p,i}^l\int_{Fo{V_2}} {{\cal N}\left( {\mathbf{x};\mathbf{x}_{p,i}^1,\mathbf{P}_{p,i}^1} \right)d\mathbf{x}} }  \le \gamma \sum\nolimits_{i = 1}^{N_p^1} {\alpha _{p,i}^1}.
\end{equation}
If, instead, the above condition is not satisfied, the GCs belonging to the subset $L_p^1$ are deleted.
An analogous procedure is followed for the unmatched subsets $L^2_p$, $p = Q+1,\ldots,M_2$ of sensor 2.
}

{
\underline{6) Fusion of the matched DIFs:}
For each association pair $ {\left( {L_p^1,L_p^2} \right)} $, $p=1,\ldots,Q$, GM approximations of the powers $[\hat v^1_p (\mathbf{x}) ]^{\omega_1}$ and $[\hat v^2_p (\mathbf{x})]^{\omega_1}$ of the corresponding DIFs
as described in Section II-C. Then, given the GMs
\begin{equation}\label{eq:GMp}
[\hat v^l_p (\mathbf{x}) ]^{\omega_l} \approx \sum\limits_{i = 1}^{{N^l_p}} {\widetilde \alpha _{p,i}^l \, {{\cal N}}(\mathbf{x}; \widetilde{\mathbf{x}}_{p,i}^l, \widetilde{\mathbf{P}}_{p,i}^l)}
\end{equation}
for $p=1,\ldots,Q$ and $l = 1,2$,  the parallelized GCI fusion structure is adopted. Accordingly, the fused PHD for the matched DIFs is obtained as
\begin{equation}\label{eq:card_mean}
\hat v^{1,2}(\mathbf{x}) = {\sum\limits_{p = 1}^Q {\sum\limits_{i = 1}^{{N_p^1}} {\sum\limits_{j = 1}^{{N_p^2}} {\alpha _{p,i,j}^{1,2} \, {\cal N}\left( {\mathbf{x};\mathbf{x}_{p,i,j}^{1,2},\mathbf{P}_{p,i,j}^{1,2}} \right)} } } }
\end{equation}
where}

{
\begin{equation}\label{eq: state_fusion_parameter}
\nonumber
\begin{aligned}
{{\mathbf{P}}}_{p,i,j}^{1,2}
&= {\left[ { \left ({\widetilde{\mathbf{P}}}_{p,i}^{1} \right)^{ - 1}} +{ \left ({\widetilde{\mathbf{P}}}_{p,j}^2 \right)^{ - 1}}\right]^{ - 1}}\\
{{\mathbf{x}}}_{i,j,m}^{1,2}
&= {{\mathbf{P}}}_{p,i,j}^{1,2} \left [ {({\widetilde{\mathbf{P}}}_{p,i}^{1})^{ - 1}}{\widetilde{\mathbf{x}}}_{p,i}^{1} + {({\widetilde{\mathbf{P}}}_{p,j}^2)^{ - 1}}{\widetilde{\mathbf{x}}}_{p,j}^2 \right]\\
\alpha _{p,i,j}^{1,2}
&= {\widetilde{\alpha}_{p,i}^1} \, {\widetilde{\alpha}_{p,j}^2} \,
{\cal N}({\widetilde{\mathbf{x}}}_{p,i}^{1} - {\widetilde{\mathbf{x}}}_{p,j}^2;0,{{{\widetilde{\mathbf{P}}}_{p,i}^1}} + {{{\widetilde{\mathbf{P}}}_{p,j}^2}}).
\end{aligned}
\end{equation}
}

{
\begin{Rem}
In terms of computational burden, the traditional GCI fusion would require to compute and store $ \left(\sum\nolimits_{p = 1}^Q {{N_p^1}}\right)  \cdot \left(\sum\nolimits_{p = 1}^Q {{N_p^2}}\right)$ GCs.
On the other hand, the parallelized GCI fusion only requires the computation of $\sum\nolimits_{p = 1}^Q \left({{N_p^1} \cdot {N_p^2}}\right)$ GCs, thus significantly reducing the computational
burden when $Q>1$.
\end{Rem}
}

{\underline{7) Fusion of the preserved DIFs:}
For each unmatched subset $L^l_p$ corresponding to a preserved DIF (as determined in step 5),
a GC approximation of the power $[\hat v^l_p (\mathbf{x}) ]^{\bar \omega_l}$  is computed as in (\ref{eq:GMp}).
Then, the total fused density takes the form
\begin{equation}\label{eq:merge_density_GM}
\begin{array}{l}
\begin{aligned}
&{{\tilde v}^{1,2}}(\mathbf{x}) = {\sum\limits_{p = 1}^Q {\sum\limits_{i = 1}^{{N_p^1}} {\sum\limits_{j = 1}^{{N_p^2}} {\alpha _{p,i,j}^{1,2} \, {{\cal N}}(\mathbf{x};\mathbf{x}_{p,i,j}^{1,2},\mathbf{P}_{p,i,j}^{1,2})} } } }\\
& + \sum\limits_{p = Q + 1}^{{M_1}} {\sum\limits_{i = 1}^{N_p^1} {{\Delta ^{1 - {{\bar \omega }_1}}} \beta _p^{12} \, \widetilde \alpha _{p,i}^1 \, {{\cal N}}\left( {\mathbf{x};\widetilde{\mathbf{x}}_{p,i}^1, \widetilde{\mathbf{P}}_{p,i}^1} \right)} } \\
& + \sum\limits_{p = Q + 1}^{{M_2}} {\sum\limits_{j = 1}^{N_p^2} {{\Delta ^{1 - {{\bar \omega }_2}}}\beta _p^{21}\, \widetilde \alpha _{p,j}^1 \, {{\cal N}}\left( {\mathbf{x}; \widetilde{\mathbf{x}}_{p,j}^2, \widetilde{\mathbf{P}}_{p,j}^2} \right)} }
\end{aligned}
\end{array}
\end{equation}
where clearly the terms corresponding to the subsets deleted at step 5 need not be computed because $\beta_p^{lr} = 0$.
}

{\underline{8) Target estimate extraction:} Following \cite{Vo2}, after the fused density has been obtained, the extraction of multi-target state estimates can be executed in a straightforward way by choosing the GCs
 whose weight $\alpha$ is greater than a given threshold, e.g., 0.5.
 }

\section{Simulation results}
\label{sec:simulation}
In this section, we test the tracking performance of the proposed fusion algorithm (CA-GCI) using the GM-PHD filter with adaptive birth model.
The proposed algorithm is compared to the C-GM-PHD filter of \cite{Vasic2016} using the OSPA error (i.e. the OSPA distance \cite{Schumacher} between the true and the estimated target sets) as performance index.
The OSPA distance parameters are set to  be  $c = 30 m$ and $p = 2$.
The errors are averaged over 200 independent Monte Carlo runs.

\emph{Tracking model and scenario}: Consider the problem of tracking an unknown and time-varying number of targets observed in clutter.
The single-target state is
\begin{equation}
\nonumber
{\mathbf{x}_k} = {[{p_{x,k}},{\dot p_{x,k}},{p_{y,k}},{\dot p_{y,k}}]^\top}
\end{equation}
where $(p_{x,k},p_{y,k})$ and $({\dot p_{x,k}},{\dot p_{y,k}})$ represent the target position and, respectively, velocity in Cartesian coordinates at time $k$.
 Each target moves according to the following dynamic model
\begin{equation}
\nonumber
{{{\mathbf{x}}} _{k + 1}} = {\mathbf{F}}{{{\mathbf{x}}} _k} + {w_k}
\end{equation}
\begin{equation}
\nonumber
\begin{aligned}
{\mathbf{F}} = \left[ {\begin{array}{*{20}{c}}
{{I_2}}\\
{{0_2}}
\end{array}\begin{array}{*{20}{c}}
{}\\
\end{array}\begin{array}{*{20}{c}}
{T_s {I_2}}\\
{{I_2}}
\end{array}} \right], {\kern 1pt} {\kern 1pt} {\kern 1pt} {\kern 1pt} {\kern 1pt} {\kern 1pt} {\kern 1pt} {\kern 1pt} {\kern 1pt} {\kern 1pt} {\kern 1pt} {\kern 1pt} {\kern 1pt} {\mathbf{Q}} = \sigma _w^2\left[ {\begin{array}{*{20}{c}}
{{\textstyle{{{T_s ^4}} \over 4}}{I_2}} \quad {{\textstyle{{{T_s ^3}} \over 2}}{I_2}}\\
{{\textstyle{{{T_s ^3}} \over 2}}{I_2}} \quad {{T_s ^2}{I_2}}
\end{array}} \right]
\end{aligned}
\end{equation}
where $I_n$ and $0_n$ denote, respectively, the $n \times n$ identity and zero matrices. $T_s=1s$ is the sampling period, $ \mathbf{Q} $ is the process noise covariance and $\sigma_w=2m/s^2$ is the standard deviation of the process noise.

\begin{figure}
\begin{center}
\includegraphics[width=0.8\columnwidth,draft=false]{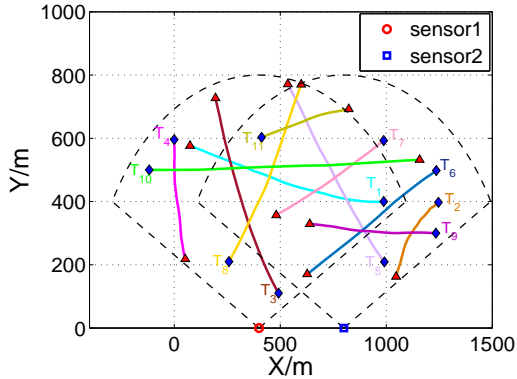}
\end{center}
\caption{True target trajectories considered in the scenario. The start/end point of each trajectory is denoted, respectively, by $\textcolor{blue}\blacklozenge|\textcolor{red}\blacktriangle$.}
\label{fig:track}
\end{figure}

The considered scenario is depicted in Fig. \ref{fig:track}. Overall, $11$ targets enter the scenario at different time instants as detailed in Table II.
Two sensors located at  $(400m,0m)$ and $(800m,0m)$, respectively, provide measurements of the unknown targets.
Each sensor has a limited FoV as depicted in Fig. \ref{fig:track}. Within the FoV the probability of detection is constant, i.e.
\begin{equation}
\nonumber
{p_{D}^l}({\mathbf{x}}) = \left\{ {\begin{array}{*{20}{c}}
{0.95,{\kern 1pt} {\kern 1pt} {\kern 1pt} {\kern 1pt} {\kern 1pt} {\mathbf{x}} \in FoV_l}\\
{0,{\kern 1pt} {\kern 1pt} {\kern 1pt} {\kern 1pt} {\kern 1pt} {\kern 1pt} {\kern 1pt} {\kern 1pt} {\kern 1pt} {\kern 1pt} {\kern 1pt} {\kern 1pt} {\kern 1pt} {\kern 1pt} {\kern 1pt} {\kern 1pt} {\kern 1pt} {\kern 1pt} {\kern 1pt} {\kern 1pt} {\mathbf{x}} \notin FoV_l}
\end{array}} \right.
\end{equation}
for $l = 1, 2$.  When a target is detected, a measurement is generated according to the linear noisy model
\begin{equation}
\begin{aligned}
\nonumber
{{{z}}_k} &= \left[ {\begin{array}{*{20}{c}}
1&0&0&0\\
0&0&1&0
\end{array}} \right]{{{\mathbf{x}}}_k}{\rm{ + }}{v_k}\\
\mathbf{R} &= \sigma_\varepsilon^2\left[ {\begin{array}{*{20}{c}}
1&0\\
0&1
\end{array}} \right]
\end{aligned}
\end{equation}
where $\mathbf{R}$ is the measurement noise covariance and $\sigma _\varepsilon=10m$ is the standard deviation of the measurement noise.
Clutter follows a uniformly distributed Poisson RFS with an average of 20 clutter points per scan ($\lambda_c=20$).

For the PHD filters, the survival probability $p_s$ is constant and equal to $0.99$. Following \cite{Houssineau2011},
the covariance of newborn targets is set to a large value so as to ensure they can cover the surveillance area as much as possible.
For the CA-GCI fusion algorithm, the pre-clustering threshold and the weight threshold are set equal to
 $T_d = 15$ and $T_{\alpha} = 0.02$, respectively. The matching threshold is $T_r = 15$.
 The fusion weights for the preserved DIFs are $\bar\omega_1 = \bar\omega_2=0.8$ and the confidence factor is $\Delta = 0.9$.
 The weights for the fusion of the matched DIFs are $\omega_1 = \omega_2=0.5$, which on average achieves nearly optimal results \cite{Uney2011}.
\begin{table}
  \centering
     \caption{Target birth and death}
    \begin{tabular}{l|l|l|l}
      \hline
      \textbf{Target} &\textbf{Init. Loc.}($m$)&\textbf{Init. Velo.}($m/s$)&\textbf{Birth/Death}($s$)\\
      \hline
      \quad$T_1$ & \quad$[1000,400]$ & \quad\quad$[-14,0]$  & \quad $1/80$ \\
      \quad$T_2$ & \quad$[1250,400]$   & \quad\quad$[-4,-2.5]$ & \quad $1/80$ \\
      \quad$T_3$ & \quad$[500,100]$ & \quad\quad$[-8,10]$ & \quad $10/60$ \\
      \quad$T_4$ & \quad$[0,600]$   & \quad\quad$[0,-4]$ & \quad $10/80$ \\
      \quad$T_5$ & \quad$[1000,200]$ & \quad\quad$[-9,9]$ & \quad $10/70$ \\
      \quad$T_6$ & \quad$[1250,505]$ & \quad\quad$[-14,-7]$  & \quad $20/60$ \\
      \quad$T_7$ & \quad$[1000,600]$   & \quad\quad$[-12,-7]$ & \quad $20/60$ \\
      \quad$T_8$ & \quad$[250,200]$ & \quad\quad$[8,10]$ & \quad $20/70$ \\
      \quad$T_9$ & \quad$[1250,300]$   & \quad\quad$[-16,0]$ & \quad $30/70$ \\
      \quad$T_{10}$ & \quad$[-150,500]$ & \quad\quad$[32,0]$ & \quad $30/70$ \\
      \quad$T_{11}$ & \quad$[400,600]$ & \quad\quad$[12,3]$ & \quad $40/80$ \\
      \hline
    \end{tabular}
\end{table}

\begin{figure}
\begin{center}
\includegraphics[width=0.8\columnwidth,draft=false]{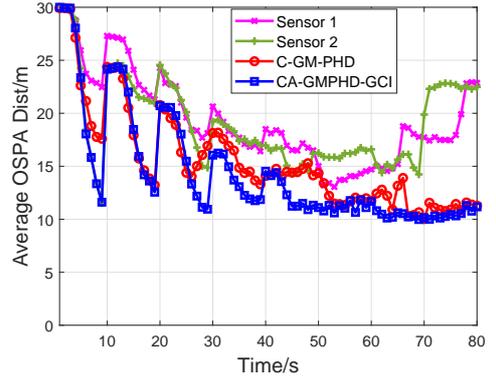}
\end{center}
\caption{Performance comparison in terms of OSPA errors among sensor 1, sensor 2, the C-GM-PHD filter and the proposed fusion algorithm (CA-GMPHD-GCI).}
\label{fig:ospa}
\end{figure}
\begin{figure}
\begin{center}
\includegraphics[width=0.8\columnwidth,draft=false]{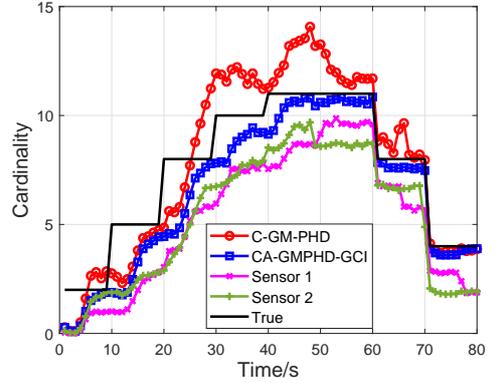}
\end{center}
\caption{Performance comparison in terms of target cardinality among sensor 1, sensor 2, the C-GM-PHD filter and the proposed fusion algorithm (CA-GMPHD-GCI).}
\label{fig:card}
\end{figure}

\emph{Simulation results}: To assess the effectiveness of the proposed algorithm, we examine the fusion performance in a situation.
Comparisons in terms of OSPA errors and cardinality estimates for the considered filters are shown in Figs. \ref{fig:ospa} and \ref{fig:card}, respectively.
From Fig. \ref{fig:ospa}, we can see that both the proposed method and the C-GM-PHD filter are able to improve the local sensor performance.
Furthermore, the proposed algorithm significantly outperforms the C-GM-PHD filter, especially in terms of cardinality estimate (Fig. \ref{fig:card}).
In fact,  the C-GM-PHD filter substantially overestimates the true cardinality during the time interval $[30, 70]s$, because the weights of the fused GCs are computed using the weights of multiple GCs.
In contrast, the proposed algorithm is able to closely follow the true cardinality, a part from a small delay in the detection of newborn targets.



In order to verify the robustness of the proposed algorithm, Monte Carlo simulations  have been performed with different detection probabilities and clutter rates. A comparison for  different $p_D$ and same $\lambda_c$ is given in
Table III. As expected for all the considered filters, the OSPA error increases as the $p_D$ increases, but the proposed algorithm provides the best performance.
Further, in Table IV a comparison for different $\lambda_c$ and  same $p_D$ is given.
Also in this case, the proposed algorithm turns out to be more robust with respect to an increase in the clutter rate.

\begin{table}[tb]
  \begin{center}
     \caption{OSPA errors with different detection probabilities and same clutter rate ($\lambda_c=20$).} \label{table4}
    \begin{tabular}{ccccc}
      \hline
      {\bf\small $P_D$} &{\bf\small Sensor 1} & {\bf\small Sensor 2}  & {\bf\small C-GM-PHD} & {\bf\small Proposed} \\
      \hline
      \hline
        {0.75} & 23.8612  & 24.0699 & 21.6949 & 21.1739\\
        \hline
        {0.85} & 21.9649  & 22.0665 & 19.4878 & 19.0956\\
        \hline
        {0.90} & 20.6366  & 20.5620 & 17.2009 & 16.5622\\
        \hline
        {0.95} & 19.4173  & 19.1632 & 15.5130 & 14.4411\\
        \hline
        {0.98} & 18.9021  & 18.3692 & 14.3381 & 13.7560\\
        \hline
    \end{tabular}
  \end{center}
\end{table}
\begin{table}[tb]
  \begin{center}
     \caption{OSPA errors with different clutter rates and same detection probability ($p_D=0.95$).} \label{table5}
    \begin{tabular}{ccccc}
      \hline
      {\bf\small $\lambda_c$} &{\bf\small Sensor 1} & {\bf\small Sensor 2}  & {\bf\small C-GM-PHD} & {\bf\small Proposed} \\
      \hline
      \hline
        {10} & 19.1457 & 18.6978 & 14.8450 & 13.3432\\
        \hline
        {20} & 19.4173 & 19.1632 & 15.5130 & 14.4411\\
        \hline
        {30} & 19.7298 & 19.5954 & 15.8714 & 15.1374\\
        \hline
        {40} & 20.2337 & 19.9615 & 16.4390 & 16.0174\\
        \hline
        {50} & 20.5805 & 20.8820 & 17.5043 & 17.1127 \\
        \hline
    \end{tabular}
  \end{center}
\end{table}

\section{Conclusions and future work}
\label{sec:conclusion}
In this paper, an effective and robust fusion algorithm has been proposed  in a distributed setting by
combining the {GCI fusion rule} with a suitable CA. Starting from an analysis of the pitfalls of the GCI fusion rule in the case of sensors with different FoVs,
a solution to this problem based on the decomposition of the local PHDs into well-separated components have been presented.
Then, two different fusion strategies have been presented: a parallelized GCI fusion strategy for combining the components in the common FoV and detected by all sensors;
and a suitable compensation strategy to preserve the components outside the common FoV and, hence, not present in all the local PHDs.
The effectiveness of the proposed approach has been analyzed  by means of simulations in a challenging tracking scenario.
An extension of the method to  the GM-CPHD filter \cite{Battistelli} considering both intensity function and cardinality distribution, is a relevant topic for future research.

\section*{Acknowledgment}
This work was supported in part by the Chang Jiang Scholars
Program, in part by the 111 Project No. B17008, in part by the National Natural
Science Foundation of China under Grant 61771110, in part by the Fundamental
Research Funds of Central Universities under Grant ZYGX2016J031, and in
part by the Chinese Postdoctoral Science Foundation under Grant 2014M550465 and Special Grant 2016T90845.

\section{Appendix}
\begin{proof}
Combining (\ref{eq:GCI_para}) with (\ref{eq:GCI_part}), we can derive the bound on ${\left\| {\hat v^{1,2}- v^{1,2}} \right\|_{{1}}}$.
For convenience, the following shorthand notation will be used
\begin{eqnarray}
\varepsilon_p^1 &= {{{\left[ {{\hat v}_p^1} \right]}^{{\omega _1}}} - {{\left[ {v_p^1} \right]}^{{\omega _1}}}}\\
\varepsilon_p^2 &= {{{\left[ {{\hat v}_p^2} \right]}^{{\omega _2}}} - {{\left[ {v_p^2} \right]}^{{\omega _2}}}}.
\end{eqnarray}
By applying the triangular inequality, the $L_1$-norm ${\left\| {{{\hat v}^{1,2}} - {v^{1,2}}} \right\|_1}$ can be decomposed into two parts as shown in (\ref{TWOparts}). Next, two parts are discussed separately.\\
{\underline{\textbf{Part 1} in (\ref{TWOparts}):}}
\begin{figure*}[htp]
\begin{equation}\label{TWOparts}
\begin{split}
&{\left\| {{{\hat v}^{1,2}} - {v^{1,2}}} \right\|_1} \\
&= {\left\| {\sum\limits_{p = 1}^Q {{{\left[ {v_p^1} \right]}^{{\omega _1}}}{{\left[ {v_p^2} \right]}^{{\omega _2}}} - \sum\limits_{p = 1}^M {{{\left[ {v_p^1} \right]}^{{\omega _1}}}{{\left[ {v_p^2} \right]}^{{\omega _2}}}} } } \right\|_1}\\
&= {\left\| {\sum\limits_{p = 1}^Q {{{\left[ {v_p^1} \right]}^{{\omega _1}}}{{\left[ {v_p^2} \right]}^{{\omega _2}}} - \left( {\sum\limits_{p = 1}^Q {{{\left[ {v_p^1} \right]}^{{\omega _1}}}{{\left[ {v_p^2} \right]}^{{\omega _2}}} + \sum\limits_{p = Q + 1}^M {{{\left[ {v_p^1} \right]}^{{\omega _1}}}{{\left[ {v_p^2} \right]}^{{\omega _2}}}} } } \right)} } \right\|_1}\\
&= {\left\| {\sum\limits_{p = 1}^Q {\left( {{{\left[ {v_p^1} \right]}^{{\omega _1}}}{{\left[ {v_p^2} \right]}^{{\omega _2}}} - {{\left[ {v_p^1} \right]}^{{\omega _1}}}{{\left[ {v_p^2} \right]}^{{\omega _2}}}} \right) + \sum\limits_{p = Q + 1}^M {{{\left[ {v_p^1} \right]}^{{\omega _1}}}{{\left[ {v_p^2} \right]}^{{\omega _2}}}} } } \right\|_1}\\
&\le \underbrace {{{\sum\limits_{p = 1}^Q {\left\| {{{\left[ {\hat{v} _p^1} \right]}^{{\omega _1}}}{{\left[ {\hat{v} _p^2} \right]}^{{\omega _2}}} - {{\left[ {v_p^1} \right]}^{{\omega _1}}}{{\left[ {v_p^2} \right]}^{{\omega _2}}}} \right\|} }_1}}_{{\textbf{part}}{\kern 1pt} {\kern 1pt} 1} + \underbrace {{{\sum\limits_{p = Q + 1}^M {\left\| {{{\left[ {v_p^1} \right]}^{{\omega _1}}}{{\left[ {v_p^2} \right]}^{{\omega _2}}}} \right\|} }_1}}_{{\textbf{part}}{\kern 1pt} {\kern 1pt} 2}
\end{split}
\end{equation}
\hrulefill 
\vspace*{1pt} 
\end{figure*}
The following upper bound can be obtained
\begin{eqnarray}
&{\kern 1pt}& \sum\limits_{p = 1}^Q {\left\| {{{\left[ {{\hat v}_p^1} \right]}^{{\omega _1}}}{{\left[ {{\hat v}_p^2} \right]}^{{\omega _2}}} - {{\left[ {v_p^1} \right]}^{{\omega _1}}}{{\left[ {v_p^2} \right]}^{{\omega _2}}}} \right\|_1}\\
\label{bound1}
&=& {\sum\limits_{p = 1}^Q {\left\| {{{\left[ {v_p^1} \right]}^{{\omega _1}}}\varepsilon _p^2 + {{\left[ {v_p^2} \right]}^{{\omega _2}}}\varepsilon _p^1 + \varepsilon _p^1\varepsilon _p^2} \right\|} _1}\\
\nonumber
&\le& \sum\limits_{p = 1}^Q {\left( {{{\left\| {{{\left[ {v_p^1} \right]}^{{\omega _1}}}\varepsilon _p^2} \right\|}_1} + {{\left\| {{{\left[ {v_p^2} \right]}^{{\omega _2}}}\varepsilon _p^1} \right\|}_1}} \right)} \\
\label{bound2}
&{\kern 1pt}& + \sum\limits_{p = 1}^Q {{{\left\| {\varepsilon _p^1\varepsilon _p^2} \right\|}_1}}\\
\nonumber
&\le& \sum\limits_{p = 1}^Q {\left( {{{\left\| {{{\left[ {v_p^1} \right]}^{{\omega _1}}}} \right\|}_1}{{\left\| {\varepsilon _p^2} \right\|}_\infty } + {{\left\| {{{\left[ {v_p^2} \right]}^{{\omega _2}}}} \right\|}_1}{{\left\| {\varepsilon _p^1} \right\|}_\infty }} \right)}\\
\label{bound3}
&{\kern 1pt}& + \sum\limits_{p = 1}^Q {{{\left\| {\varepsilon _p^1} \right\|}_1}{{\left\| {\varepsilon _p^2} \right\|}_\infty }}\\
\nonumber
&\le& \sum\limits_{p = 1}^Q {\left( {{{\left\| {{{\left[ {v_p^1} \right]}^{{\omega _1}}}} \right\|}_1}{{\left\| {\varepsilon _p^2} \right\|}_\infty } + {{\left\| {{{\left[ {v_p^2} \right]}^{{\omega _2}}}} \right\|}_1}{{\left\| {\varepsilon _p^1} \right\|}_\infty }} \right)} \\
\label{bound4}
&{\kern 1pt}& + \sum\limits_{p = 1}^Q {\left( {{{\left\| {{{\left[ {{\hat v}_p^1} \right]}^{{\omega _1}}}} \right\|}_1} + {{\left\| {{{\left[ {v_p^1} \right]}^{{\omega _1}}}} \right\|}_1}} \right){{\left\| {\varepsilon _p^2} \right\|}_\infty }}\\
\label{bound5}
&\le& QK\left( {3{\delta ^{{\omega _2}}} + {\delta ^{{\omega _1}}}} \right)
\end{eqnarray}
where
\begin{enumerate}
  \item (\ref{bound1}) holds because
      \begin{equation}
      \nonumber
      \begin{split}
      ab - cd &= (a - c + c)(b - d + d) - cd\\
      &= (a - c)(b - d) + d(a - c) + c(b - d).
     \end{split}
     \end{equation}
  \item (\ref{bound2}) is obtained by applying the triangular inequality;
  \item (\ref{bound3}) holds because $\| f \, g \|_1 \le \| f \|_\infty  \, \| g \|_1$;
  \item (\ref{bound4}) can be ontained in a similar way as (\ref{bound1});
  \item As to (\ref{bound5}),  in view of (\ref{bounding})  and (\ref{bounded}), we have
      \begin{eqnarray}
      \label{proof1}
      {\left\| {\hat v_m^1 - v_m^1} \right\|_\infty } &\le& \delta \\
      \label{proof2}
      {\left\| {\hat v_m^2 - v_m^2} \right\|_\infty } &\le& \delta \\
      \label{proof3}
      {{{\left\| {{{\left[ {v_p^1} \right]}^{{\omega _1}}}} \right\|}_1}} &\le& K \\
      \label{proof4}
      {{{\left\| {{{\left[ {v_p^2} \right]}^{{\omega _2}}}} \right\|}_1}} &\le& K \\
      \label{proof5}
      {{{\left\| {{{\left[ {{\hat v}_p^1} \right]}^{{\omega _1}}}} \right\|}_1}} &\le& K
      \end{eqnarray}
Then, since $\omega_1 \le 1$ and $\omega_2 \le 1$, we have
\begin{eqnarray}
{\left\| {{{\left[ {\hat v_p^1} \right]}^{{\omega _1}}} - {{\left[ {v_p^1} \right]}^{{\omega _1}}}} \right\|_\infty } \le \delta^{\omega_1}\\
 {\left\| {{{\left[ {\hat v_p^2} \right]}^{{\omega _2}}} - {{\left[ {v_p^2} \right]}^{{\omega _2}}}} \right\|_\infty } \le \delta^{\omega_2}.
\end{eqnarray}
As a result, (\ref{bound5}) is obtained.
\end{enumerate}

{\underline{\textbf{Part 2} in (\ref{TWOparts}):}}
Similarly to part 1, the following upper bound can be obtained

\begin{eqnarray}
\nonumber
&{\kern 1pt}&{{{\sum\limits_{p = Q + 1}^M {\left\| {{{\left[ {v_p^1} \right]}^{{\omega _1}}}{{\left[ {v_p^2} \right]}^{{\omega _2}}}} \right\|}_1 }}} \\
\nonumber
&=& {\sum\limits_{p = Q + 1}^{M_1} {\left\| {{{\left[ {v_p^1} \right]}^{{\omega _1}}}{{\left[ {v_p^2} \right]}^{{\omega _2}}}} \right\|} _1} + {\sum\limits_{p = M_1 + 1}^M {\left\| {{{\left[ {v_p^1} \right]}^{{\omega _1}}}{{\left[ {v_p^2} \right]}^{{\omega _2}}}} \right\|} _1} \\
\nonumber
&\le& {\sum\limits_{p = Q + 1}^{{M_1}} {\left\| {{{\left[ {v_p^1} \right]}^{{\omega _1}}}} \right\|} _1}{\left\| {{{\left[ {v_p^2} \right]}^{{\omega _2}}}} \right\|_\infty } \\
\label{enlarge1_part2}
&\quad& + {\sum\limits_{p = {M_1} + 1}^M {\left\| {{{\left[ {v_p^2} \right]}^{{\omega _2}}}} \right\|} _1}{\left\| {{{\left[ {v_p^1} \right]}^{{\omega _1}}}} \right\|_\infty }\\
&=& {\sum\limits_{p= Q + 1}^{{M_1}} {{\delta ^{{\omega _2}}}\left\| {{{\left[ {v_p^1} \right]}^{{\omega _1}}}} \right\|} _1} + {\sum\limits_{p = {M_1} + 1}^M {{\delta ^{{\omega _1}}}\left\| {{{\left[ {v_p^2} \right]}^{{\omega _2}}}} \right\|} _1}\\
\label{enlarge2_part2}
&\le& \left( {{M_1} - Q} \right)K{\delta ^{{\omega _2}}} + \left( {M - {M_1}} \right)K{\delta ^{{\omega _1}}}.
\end{eqnarray}
Combining (\ref{bound5}) and (\ref{enlarge2_part2}) and recalling that $M = M_1+M_2-Q$, the upper bound
\begin{equation}
{\left\| {\hat v^{1,2} - v^{1,2}} \right\|_{{1}}} \le {M_2}K{\delta ^{{\omega _1}}} + \left( {2Q + {M_1}} \right)K{\delta ^{{\omega _2}}}
\end{equation}
can be readily obtained. Hence, Proposition 1 holds.
\end{proof}

\end{document}